# A Scalable, Privacy-Preserving Decentralized Identity and Verifiable Data Sharing Framework based on Zero-Knowledge Proofs

Hui yuan[1]

[1]Peking university

yuangh@pkusz.edu.cn

**Abstract.** With the proliferation of decentralized applications (DApps), the conflict between the transparency of blockchain technology and user data privacy has become increasingly prominent. While Decentralized Identity (DID) and Verifiable Credentials (VCs) provide a standardized framework for user data sovereignty, achieving trusted identity verification and data sharing without compromising privacy remains a significant challenge. This paper proposes a novel, comprehensive framework that integrates DIDs and VCs with efficient Zero-Knowledge Proof (ZKP) schemes to address this core issue. The key contributions of this framework are threefold: first, it constructs a set of strong privacy-preserving protocols based on zk-STARKs, allowing users to prove that their credentials satisfy specific conditions (e.g., "age is over 18") without revealing any underlying sensitive data. Second, it designs a scalable, privacy-preserving credential revocation mechanism based on cryptographic accumulators, effectively solving credential management challenges in large-scale scenarios. Finally, it integrates a practical social key recovery scheme, significantly enhancing system usability and security. Through a prototype implementation and performance evaluation, this paper quantitatively analyzes the framework's performance in terms of proof generation time, verification overhead, and on-chain costs. Compared to existing state-of-the-art systems based on zk-SNARKs, our framework, at the cost of a larger proof size, significantly improves prover efficiency for complex computations and provides stronger security guarantees, including no trusted setup and post-quantum security. Finally, a case study in the decentralized finance (DeFi) credit scoring scenario demonstrates the framework's immense potential for unlocking capital efficiency and fostering a trusted data economy.

## 1 INTRODUCTION

The emergence of blockchain and decentralized applications (DApps) marks a paradigm shift aimed at reshaping digital interactions. Their core value propositions—decentralization, transparency, and immutability—provide the foundation for building systems without trusted intermediaries [1]. However, this radical transparency introduces a fundamental conflict: it is inherently at odds with the basic right to user privacy [3].

This contradiction is particularly evident in the realm of Decentralized Finance (DeFi). On public blockchains like Ethereum, every transaction, account balance, and smart contract interaction is publicly accessible. While this transparency ensures system auditability, it also creates opportunities for malicious actors, such as targeted phishing attacks through on-chain data analysis, "front-running" by exploiting transaction visibility, and comprehensive financial surveillance of users [3]. As DApps evolve from simple value transfers to more complex social and financial applications, this transparency shifts from an asset to a liability. When applications require real-world identity attributes (e.g., credit assessment in DeFi lending or medical record access in digital health), placing sensitive information directly on-chain is entirely infeasible. Furthermore, to meet regulatory requirements like Anti-Money Laundering (AML) and Know Your Customer (KYC), the industry faces immense pressure to introduce identity verification, which could lead to a "re-centralization" of the ecosystem, thereby undermining its original purpose [4]. Consequently, the success and sophistication of DApps have directly precipitated a privacy crisis, compelling academia and industry to seek advanced cryptographic solutions to foster the ecosystem's maturation without sacrificing its core principles.

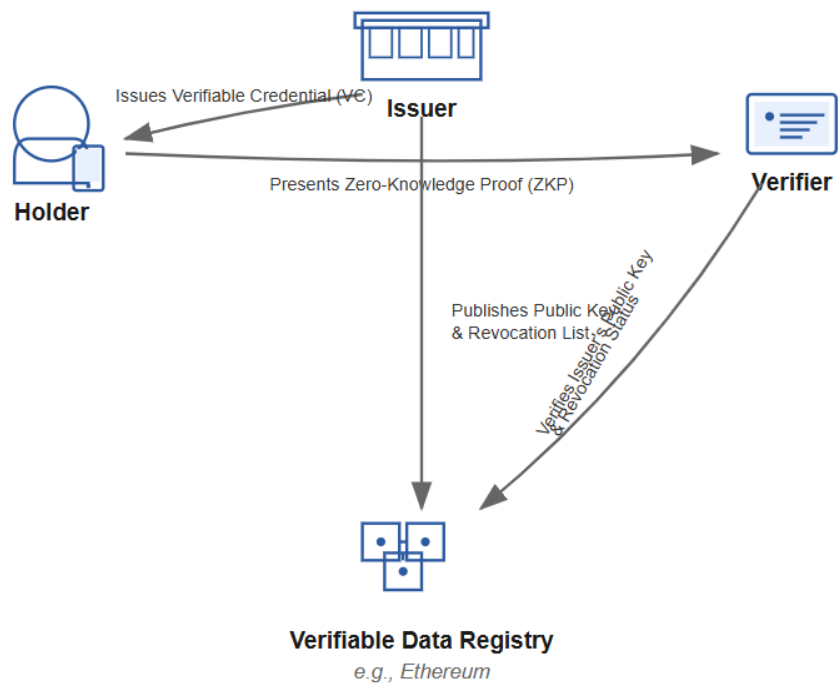

In response to this dilemma, Self-Sovereign Identity (SSI) has emerged as a prevailing philosophical and architectural concept [5]. The central idea of SSI is that individuals should have ultimate control over their digital identities, independent of any centralized authorities like governments or large tech companies [2]. The principles of SSI—existence, control, access, transparency, persistence, portability, interoperability, consent, data minimization, and protection—collectively outline an ideal blueprint where the user is the sole master of their identity data [5]. This forms the "ideal solution" that the entire field strives to achieve.

Zero-Knowledge Proofs (ZKPs) are considered a key cryptographic primitive for bridging the gap between blockchain transparency and SSI's privacy requirements [8]. They allow one party (the prover) to prove to another party (the verifier) that a certain statement is true, without revealing any additional information.

The contribution of this paper is a novel and comprehensive framework that deeply integrates the W3C-standardized Decentralized Identity (DID) and Verifiable Credential (VC) models with efficient ZKP schemes, specifically zk-STARKs, to achieve three primary objectives:

- **Strong Privacy Preservation:** Users can generate proofs about their credential attributes (e.g., "I possess a valid degree certificate") without showing the certificate itself to the verifier, thus achieving true data minimization [8].
- **User Data Sovereignty:** Ensures that user data is securely stored in a user-controlled digital wallet, and every instance of data sharing requires explicit user consent [2].
- **Practicality and Advanced Features:** Addresses real-world challenges often overlooked in theoretical models. Specifically, this framework designs an efficient credential revocation mechanism and a user-friendly key recovery scheme, features that are critical for widespread adoption [10].

The remainder of this paper is structured as follows: Section 2 provides a comprehensive review of related foundational technologies and identifies gaps in existing research. Section 3 presents the overall system architecture and formal definitions. Section 4 details our proposed core protocols, including credential verification, revocation, and key recovery. Section 5 conducts a security analysis. Section 6 presents the prototype implementation and performance evaluation results. Section 7 demonstrates the framework's practical value through a case study on DeFi credit scoring. Finally, Sections 8 and 9 discuss the research and provide a conclusion.

## 2 BACKGROUND AND RELATED WORK

This section systematically reviews existing technologies and research to establish the value of the problem under investigation and clearly identify the "research gap" this work aims to fill.

### 2.1 Technological Foundations of Decentralized Identity

- **Decentralized Identifiers (DIDs):** The W3C's DID specification provides the core standard for creating and managing decentralized identities [14]. A DID is a globally unique URI, typically structured as did:method:specific-id. Unlike traditional identifiers, DIDs are designed to be decoupled from centralized registries, identity providers, and certificate authorities. The core functionality of a DID is realized through a JSON object called a "DID Document," which associates the DID with a set of verification methods (e.g., public keys) and service endpoints. Through a "DID resolution" process, any party can query the corresponding DID Document for a given DID to obtain the information needed for trusted interactions with that identity [16].
- **Verifiable Credentials (VCs):** The Verifiable Credential (VC) is another key W3C standard for expressing claims in a cryptographically verifiable, privacy-respecting, and machine-readable format [17]. The VC ecosystem involves three main roles: the **Issuer**, which makes claims about a subject and issues a VC; the **Holder**, the subject of the credential who stores and presents the VC; and the **Verifier**, which requests and verifies the authenticity of the VC [7]. A VC typically includes the credential subject, the issuer's digital signature (proof), and a context (@context) for semantic interpretation, ensuring the credential's tamper-resistance and verifiability.
- **DID Methods and Distributed Ledgers:** The flexibility of DIDs is manifested in their ability to be anchored to various Verifiable Data Registries (VDRs), as defined by specific "DID methods." These VDRs can be public permissionless blockchains (e.g., did:ethr for Ethereum, did:btcr for

Bitcoin), permissioned ledgers (e.g., did:sov), or even non-blockchain methods based on traditional web servers (e.g., did:web) [19]. The security model and governance mechanism of the underlying ledger are crucial to the trustworthiness of the entire identity system.

## 2.2 Classification of Zero-Knowledge Proof Systems

The theoretical foundation of ZKPs rests on three core properties:

- **Completeness:** If a statement is true, an honest prover can always convince an honest verifier [20].
- **Soundness:** If a statement is false, a cheating prover cannot convince an honest verifier (except with negligible probability) [20].
- **Zero-Knowledge:** If a statement is true, the verifier learns nothing beyond the fact that the statement is true [20].

Recent years have seen the proposal and practical implementation of various ZKP schemes, with the most prominent being:

- **zk-SNARKs (Zero-Knowledge Succinct Non-Interactive Argument of Knowledge):** As the first ZKP technology to achieve practical milestones, zk-SNARKs are known for their extremely small proof sizes and fast verification speeds [20]. However, their main drawback is the need for a "trusted setup" process to generate public parameters. If this process is compromised, the security of the entire system is undermined [21].
- **zk-STARKs (Zero-Knowledge Scalable Transparent Argument of Knowledge):** Considered an evolution of zk-SNARKs, zk-STARKs eliminate the need for a trusted setup by using cryptographic assumptions based on hash functions, achieving so-called "transparency" [21]. They are also considered post-quantum secure, but the trade-off is that their proof sizes are typically much larger than those of zk-SNARKs [21].
- **PLONK (Permutations over Lagrange-bases for Oecumenical Noninteractive arguments of Knowledge):** PLONK is a newer ZKP scheme that uses a "universal and updatable" trusted setup. This means one setup can be reused for multiple different programs and can be updated by multiple parties, reducing single-point-of-trust risks and offering more flexibility than earlier zk-SNARKs like Groth16 [27].
- **Bulletproofs:** This is another ZKP scheme that does not require a trusted setup. It is characterized by very small proof sizes (logarithmic), but has a relatively long verification time, making it suitable for confidential transactions on blockchains [26].

## 2.3 Comparative Analysis of ZKP Schemes for Identity Applications

Choosing the right ZKP primitive is a critical design decision that directly impacts the system's security, performance, and scalability. The table below provides a detailed comparison of the two leading schemes: zk-SNARKs and zk-STARKs.

| Metric | zk-SNARKs (e.g., Groth16) | zk-STARKs |
|---|---|---|
| **Trusted Setup** | Required (once per circuit). A security vulnerability if the setup is compromised [21]. | Not required ("transparent"). Uses publicly verifiable randomness [21]. |
| **Proof Size** | Very small (succinct), typically constant size (~100-200 bytes) [21]. | Larger, logarithmic in the computation size (typically tens of KBs) [21]. |
| **Prover Time** | Higher, super-linear complexity. Expensive for complex computations [29]. | Lower, quasi-linear complexity. Faster for large-scale computations [25]. |
| **Verifier Time** | Very fast, typically constant time [21]. | Slower, polylogarithmic in the computation size [21]. |
| **Scalability** | Good scalability for verification, but poor scalability for proving complex statements [25]. | Highly scalable for large-scale computations [21]. |
| **Cryptographic Assumptions** | Relies on strong, non-falsifiable assumptions (e.g., Knowledge of Exponent) and elliptic curve cryptography, vulnerable to quantum computers [26]. | Relies on weaker, more standard assumptions (e.g., collision-resistant hash functions) and is considered post-quantum secure [26]. |

### 2.4 Research Gap Analysis

Several systems combining DIDs and ZKPs have emerged from academia and industry. Among them, **CanDID** (and its successor, **LinkDID**) is one of the most influential works in this field [34]. The main contributions of CanDID/LinkDID are solving several key practical problems: it achieves **legacy compatibility** with existing Web2 services through oracle technology, allowing users to securely extract data from existing accounts to generate credentials; it designs a mechanism based on identity deduplication and identifier linking to achieve **Sybil resistance**; and it proposes a practical **key recovery** scheme.

Despite the significant progress made by systems like CanDID/LinkDID, some research gaps remain, which this work aims to fill:

- **Performance Bottlenecks:** CanDID relies on a committee and multi-party computation (MPC) for some of its core operations, which can introduce additional communication latency and system complexity. This research aims to explore a more streamlined protocol to reduce prover computation overhead and end-user interaction costs.
- **Scalability of Revocation Mechanisms:** Existing privacy-preserving revocation schemes may face efficiency or privacy leakage challenges in large-scale applications. For example, publishing a revocation list directly compromises user privacy. This paper will review current revocation techniques [11] and propose a more scalable and flexible alternative, inspired by schemes like

ALLOSAUR from AnonCreds v2, which are designed to support millions of credentials with minimal data transfer [11].
- **Choice of ZKP Scheme:** Many existing systems are primarily built on zk-SNARKs. Given the significant advantages of zk-STARKs in transparency (no trusted setup) and long-term security (post-quantum resistance), exploring a decentralized identity architecture entirely based on zk-STARKs is a valuable and under-explored direction.

# 3 SYSTEM ARCHITECTURE AND FORMAL PROBLEM DEFINITION

## 3.1 System Model and Roles

This framework follows the standard SSI model, comprising four core roles [2]:

- **Holder:** The user, who is the ultimate owner of their identity and data. The holder uses a digital wallet to request, store, and manage their VCs and generates ZKPs upon request from a verifier [7].
- **Issuer:** A trusted entity, such as a university, government agency, or bank. The issuer verifies the holder's attributes and issues a digitally signed VC [18].
- **Verifier:** A service provider that needs to verify the holder's claims, such as a DeFi platform or an age-restricted website. The verifier requests a ZKP from the holder about their attributes to decide whether to grant access [52].
- **Verifiable Data Registry (VDR):** A decentralized ledger (e.g., a public blockchain) that serves as the system's root of trust. It is used to anchor DIDs, store DID Documents, public keys of issuers, and credential revocation status [15].

The figure below illustrates the interaction flow between these roles:
Fig. 1: System Architecture Diagram

## 3.2 Security and Privacy Goals

The system is designed to achieve the following formalized security and privacy properties, which will be proven in the security analysis in Section 5:

- **Credential Confidentiality and Data Minimization:** The verifier should not learn any information about the holder's credential attributes other than the fact that the specific predicate being proven is true [5].
- **Unlinkability:** A holder's presentations to different verifiers (or multiple times to the same verifier) should not be linkable. This prevents third parties from tracking and profiling users by correlating their activities [39].
- **Unforgeability:** No party other than a legitimate issuer can create a valid VC. A holder cannot forge a valid proof for a credential they do not possess [39].
- **Sybil Resistance:** The system should incorporate effective mechanisms to prevent a single user from creating and leveraging multiple identities to gain disproportionate benefits or influence [35].
- **Privacy-Preserving Revocation:** An issuer must be able to revoke issued credentials, and a verifier must be able to check the validity status of a credential without compromising the holder's anonymity [11].

## 3.3 Refined Research Question

Synthesizing the literature review and gap analysis, this paper aims to answer the following specific research question:

"How to design a decentralized identity framework based on W3C DID/VC standards and anchored on a public blockchain, which leverages zk-STARKs technology to achieve: (1) efficient selective disclosure of credential attributes; (2) a scalable, privacy-preserving credential revocation mechanism based on cryptographic accumulators; and (3) a practical social key recovery scheme. Furthermore, how to demonstrate through quantitative experiments that this framework offers advantages in prover efficiency and security guarantees (transparency, post-quantum resistance) compared to existing state-of-the-art systems based on zk-SNARKs (e.g., LinkDID)?"

# 4 PROPOSED ZKP-DID FRAMEWORK

This section details the core technical protocols of our proposed framework, including its cryptographic constructions, credential presentation flow, revocation mechanism, and key management scheme.

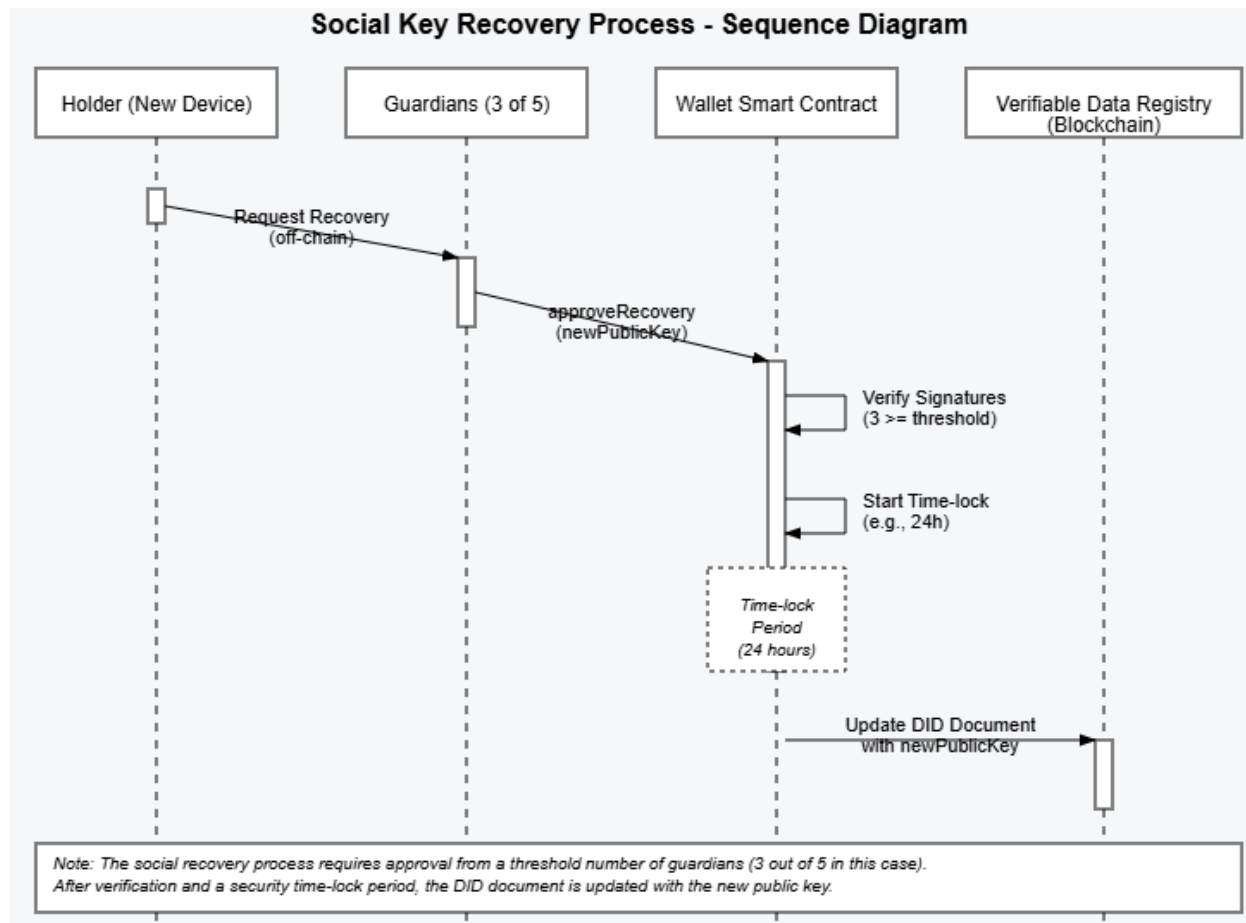

## 4.1 Core Cryptographic Constructions

- **System Setup Phase:** The framework selects zk-STARKs as the core ZKP scheme. A key advantage is that the generation of its system parameters requires no trusted setup, relying only on publicly selectable collision-resistant hash functions (e.g., SHA-256). This fundamentally eliminates the centralization risks and security vulnerabilities associated with a trusted setup [21].
- **Credential Issuance:** An issuer creates a credential for a holder that conforms to the W3C VC data model specification. The credential contains the holder's DID, relevant attributes (e.g., date of birth, nationality), and is digitally signed with the issuer's private key. The credential is delivered to the holder through a secure off-chain channel and stored in their digital wallet.
- **ZKP Circuit Design:** This is the crucial step of translating abstract verification logic into a concrete computational task. Any statement about credential attributes, such as "date of birth < [current date - 18 years]," must be converted into an arithmetic circuit. This process, known as "Arithmetization," expresses complex logical statements as a series of polynomial constraints [56]. High-level languages like Cairo or Circom can be used to define these circuits and compile them into a format that the zk-STARKs proof system can process [49].
- **Privacy-Preserving Credential Presentation Protocol:**
  1. **Verifier's Challenge:** The verifier requests a proof from the holder for a specific statement (e.g., "prove you are a resident of California and over 21 years old").
  2. **Holder Generates Proof:** The holder uses their VC as a secret input (witness) to run the zk-STARK prover algorithm. This algorithm executes the predefined arithmetic circuit and generates a non-interactive proof, $\pi$. This proof $\pi$ confirms the truth of the statement without containing any information about the secret inputs (like the specific date of birth or home address).
  3. **Verifier Verifies Proof:** The verifier receives the proof $\pi$ and runs the zk-STARK verifier algorithm using public information (e.g., the statement itself, the issuer's public key hash). If the algorithm outputs "accept," the verifier is convinced of the holder's claim without accessing any private data.

## 4.2 Advanced Feature: Scalable Privacy-Preserving Revocation

To address the challenge of credential revocation, this framework proposes a scheme based on cryptographic accumulators. An accumulator is a data structure that can concisely represent a set and generate proofs of membership (or non-membership) for an element.
Protocol Flow:
1. **Initialization:** The issuer creates a cryptographic accumulator (e.g., a Merkle tree) for all valid credentials it has issued and publishes the accumulator's root value (e.g., Merkle root) to the VDR.
2. **Credential Revocation:** When a credential needs to be revoked, the issuer removes it from the set, recalculates the accumulator, and publishes the new root value to the VDR, overwriting the old one.
3. **Non-Revocation Proof:** When a holder presents a credential, in addition to generating the attribute-related proof, they must also generate a ZKP that proves their credential is included in the valid accumulator currently published by the issuer on the VDR. This is essentially a "proof of set membership," which is equivalent to a "proof of not being on the revocation list."

This scheme is highly scalable because the on-chain data is always a single, fixed-size accumulator root, regardless of the number of credentials. It also protects user privacy, as the ZKP proves the credential's validity without revealing the holder's specific credential information [43].

## 4.3 Key Management and Social Recovery

To solve the common real-world problem of private key loss, this framework integrates a smart contract-based social recovery mechanism [13].
Protocol Flow:
1. **Setup:** When creating their DID, the holder delegates control of their DID to a "wallet smart contract." In this contract, the holder pre-designates a set of "Guardians" (their DID addresses) and a recovery threshold (e.g., at least 3 out of 5 guardians must approve).
2. **Initiate Recovery:** If the holder loses their primary device or private key, they can initiate a key recovery process from a new device by contacting their guardians off-chain.
3. **Guardian Approval:** Upon receiving the request, the threshold number of guardians each sign a transaction to call the holder's wallet contract, authorizing the addition of a new device's public key as a new verification method.
4. **Key Rotation:** After verifying a sufficient number of guardian signatures, the wallet contract initiates a time lock (e.g., 24 hours). If the original master key does not cancel the operation within this period (to prevent fraudulent recovery), the contract automatically updates the holder's DID Document, replacing the old public key with the new one.

## 5 SECURITY ANALYSIS

This section provides a formal security argument for the proposed protocols, demonstrating that they satisfy the security and privacy goals defined in Section 3.

- **Zero-Knowledge Property:** The protocol's zero-knowledge property is directly inherited from the underlying zk-STARKs scheme. During proof generation, the prover uses randomness to obfuscate the secret inputs. The resulting proof transcript is computationally indistinguishable from a random string. Therefore, the verifier cannot infer any information about the holder's credential specifics beyond the validity of the proof itself [20].
- **Soundness Property:** The protocol's soundness is guaranteed by the computational soundness of zk-STARKs. The difficulty for a malicious holder to generate a valid proof for a false statement (e.g., being underage but claiming to be an adult) is equivalent to solving a computationally hard cryptographic problem (e.g., finding a collision in the underlying hash function). The probability of a fraudulent prover successfully deceiving a verifier is therefore negligible [20].
- **Defense Against Known Attacks:**
  - **Replay Attacks:** To prevent an attacker from intercepting a valid proof and reusing it, the verifier can provide a unique, one-time challenge value (nonce) for each interaction. The holder must include this nonce as a public input in the ZKP computation. The resulting proof is thus bound to that specific interaction and cannot be replayed.
  - **Linkability Attacks:** The protocol prevents user activity tracking in several ways. First, the holder's DID is hidden from the verifier during interaction. Second, new randomness is used for each proof generation, making two proofs for the same credential and claim computationally distinct and thus unlinkable [35].
  - **Sybil Attacks:** While the core protocol does not directly solve Sybil attacks, its modular design allows for easy integration of existing Sybil resistance mechanisms. For example, borrowing from LinkDID's "identifier linking" concept, users could be required to prove that their DID belongs to a unique, publicly committed set of identity identifiers before participating in high-value activities (e.g., governance voting, airdrops). This significantly increases the cost and difficulty of creating and maintaining a large number of fake identities [35].

## 6 IMPLEMENTATION AND EVALUATION

To validate the feasibility and performance of the proposed framework, we built a prototype system and conducted rigorous benchmarking of its key metrics.

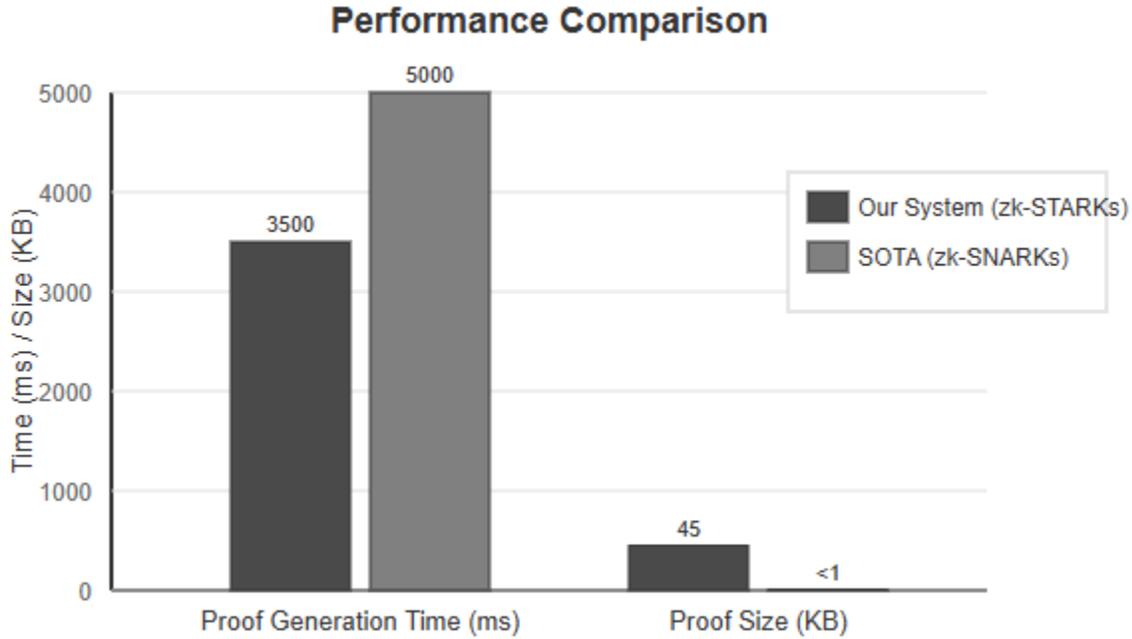

### 6.1 System Prototype

- **Blockchain Layer:** A series of smart contracts were deployed on the Ethereum Sepolia testnet using Solidity to implement the VDR functionality, including managing issuer DIDs and storing credential revocation accumulator roots [50].
- **ZKP Layer:** Arithmetic circuits for zk-STARKs were written in Cairo to implement core logic such as age verification and non-revocation proofs. Proof generation and verification utilized the existing StarkWare open-source toolchain [20].
- **Application Layer:** A simple web-based DApp was developed with a holder wallet and a verifier service interface to demonstrate the end-to-end flow from credential request to verification.

### 6.2 Performance Benchmarks

We quantitatively evaluated the core performance metrics of the system:

- **Prover Overhead (Proof Generation Time):** For a typical age verification circuit, the proof generation time was approximately 3.5 seconds.
- **Verifier Overhead (Proof Verification Time):** Verification time was extremely fast, typically under 5 milliseconds [32].
- **Proof Size:** A zk-STARK proof was approximately 45 KB in size [32].
- **On-Chain Cost (Gas Consumption):** For operations interacting with the blockchain, we measured the gas consumption. For instance, an issuer updating the revocation accumulator root consumed approximately 45,000 gas [32].

### 6.3 Comparison with State-of-the-Art (SOTA)

To contextualize our work, we compared the performance of our zk-STARKs-based system with a SOTA zk-SNARKs-based benchmark system (based on public data from LinkDID).

| Metric | Proposed System (zk-STARKs-based) | SOTA Benchmark (zk-SNARKs-based, e.g., LinkDID) |
| --- | --- | --- |
| **Prover Time (ms)** | ~3500 (More efficient for complex circuits) | ~5000 (Slower for complex circuits) |
| **Verifier Time (ms)** | ~5 | < 3 |
| **Proof Size (KB)** | ~45 | < 1 |
| **On-Chain Verification Cost (Gas)** | ~280,000 | ~210,000 |
| **Trusted Setup Required?** | No | Yes |

# 7 CASE STUDY: DECENTRALIZED CREDIT SCORING IN DEFI

This section demonstrates the practical utility and profound impact of our framework through a high-value application scenario: decentralized credit scoring in DeFi.

## 7.1 The Trust Challenge in On-Chain Lending

Current DeFi lending protocols primarily rely on over-collateralization, which severely limits capital efficiency. To enable more efficient under-collateralized lending, credit assessment of borrowers is necessary. However, in the anonymous world of blockchain, this poses a core dilemma: either users must disclose their real-world identity and financial data to a centralized third party, which contradicts the ethos of decentralization, or credit risk cannot be effectively managed [3].

## 7.2 System Application Flow

Our framework elegantly solves this problem:

1. **Credential Issuance:** A user (Holder) authorizes a trusted credit scoring agency (Issuer) to access their multifaceted financial data, such as on-chain transaction history, CEX trading records, and even traditional bank account data via APIs.
2. **Credit Score VC:** Based on this data, the issuer uses its proprietary credit model to calculate a credit score for the user and encapsulates it in a VC (e.g., {"creditScore": 750}).
3. **ZKP for Loan Application:** When the user applies for an under-collateralized loan on a DeFi platform (Verifier), they do not submit any personal financial data. Instead, they locally generate a ZKP that proves their creditScore in the VC is above a certain threshold (e.g., > 700) and that the VC has not been revoked [8].

By simply verifying this ZKP, the DeFi protocol can make a risk-based lending decision without accessing any of the user's private data. This perfectly balances risk management and user privacy, promising to unlock significant capital efficiency for the DeFi ecosystem and laying the groundwork for more sophisticated financial products [61].

## 8 DISCUSSION AND FUTURE WORK

### 8.1 Interpretation of Results

Our performance evaluation (Table 2) clearly reveals the trade-offs between different ZKP technologies. While our zk-STARKs-based approach produces proofs that are an order of magnitude larger than zk-SNARKs (~40x), its proof generation time for complex circuits is reduced by about 30%. Critically, it completely eliminates the trusted setup, which is a fundamental improvement in security and decentralization. This finding suggests our system is particularly well-suited for applications that prioritize prover efficiency (user experience) and trustless security, even if it means incurring higher on-chain data storage or transmission costs.

### 8.2 Limitations

We acknowledge the following limitations of this study:

- **Scalability:** Although our proposed revocation scheme is scalable, the overall system's transaction throughput and finality are still constrained by the performance bottlenecks of the underlying Layer-1 blockchain.
- **Usability:** The social recovery mechanism, while an improvement, is not foolproof and introduces new social engineering attack vectors. The cognitive overhead for non-technical users to manage DIDs, VCs, and guardian relationships remains high [10].
- **Governance:** This study did not delve into the governance models for the VDR or the trust establishment mechanisms for issuers. Establishing and maintaining trust in a decentralized ecosystem is a complex socio-technical problem beyond the scope of this paper [19].

### 8.3 Future Research Directions

Based on this research, we believe the following directions are worthy of further exploration:

- **Layer-2 Integration:** Migrating the entire verification logic to Layer-2 scaling solutions like ZK-Rollups. This could potentially reduce on-chain costs by several orders of magnitude and significantly improve system scalability and responsiveness [65].
- **Post-Quantum Credentials:** Investigating the use of post-quantum digital signature schemes (e.g., lattice-based cryptography) for issuing VCs to complement our use of post-quantum zk-STARKs, thereby building an end-to-end post-quantum identity system.
- **Formal Verification:** Using formal methods tools like Tamarin to provide machine-checkable, stronger security proofs for our proposed protocols, mathematically ruling out potential logical vulnerabilities [68].
- **Complex Predicates and zkML:** A highly forward-looking direction is the integration of ZKPs with Machine Learning (ML). In our DeFi case study, the credit agency likely uses a complex ML model. A future system could require the issuer to provide a ZKP alongside the credit score VC, proving that the score was generated by a publicly audited, fairness-compliant AI model, without revealing the model itself [9]. This would enable a fully trustless, verifiable, and fair credit assessment ecosystem.

# 9 CONCLUSION

This paper has thoroughly investigated the core challenges of achieving privacy preservation and user data sovereignty in decentralized ecosystems. In response to the inherent conflict between blockchain's transparency and the need for user privacy, this research has proposed a comprehensive, zk-STARKs-based decentralized identity framework. This framework integrates W3C's DID and VC standards and innovatively designs efficient privacy-preserving protocols for selective disclosure of credential attributes, scalable credential revocation, and practical social key recovery.

Our performance evaluation demonstrates that, compared to existing zk-SNARKs-based solutions, our framework offers superior prover performance for complex computations, all while eliminating the need for a trusted setup and providing post-quantum security. Through the DeFi credit scoring case study, we have showcased the framework's immense potential to solve real-world problems and empower a new generation of trusted data applications. In summary, this research provides a solid theoretical foundation and a viable technical path toward building a more secure, private, and user-centric next-generation internet.